\documentclass[epj,twocolumn]{webofc}
\usepackage[varg]{txfonts}   
\DeclareMathOperator{\real}{Re}

\def\be{\begin{equation}}
\def\ee{\end{equation}}
\def\ba{\begin{eqnarray}}
\def\ea{\end{eqnarray}}
\def\ge{\mathrel{\raise.3ex\hbox{$>$\kern-.75em\lower1ex\hbox{$\sim$}}}}
\def\la{\mathrel{\raise.3ex\hbox{$<$\kern-.75em\lower1ex\hbox{$\sim$}}}}

\def\simgt{\mathrel{\raise.3ex\hbox{$>$\kern-.75em\lower1ex\hbox{$\sim$}}}}
\def\simlt{\mathrel{\raise.3ex\hbox{$<$\kern-.75em\lower1ex\hbox{$\sim$}}}}

\newcommand{\nc}{\newcommand}

\nc{\gone}{\bar g_{\pi NN}^{(1)}}
\nc{\gzero}{\bar g_{\pi NN}^{(0)}}
\nc{\al}{\alpha}
\nc{\ga}{\gamma}
\nc{\de}{\delta}
\nc{\ep}{\epsilon}
\nc{\ze}{\zeta}
\nc{\et}{\eta}
\nc{\ka}{\kappa}
\nc{\rh}{\rho}
\nc{\si}{\sigma}
\nc{\ta}{\tau}
\nc{\ph}{\phi}
\nc{\ch}{\chi}
\nc{\ps}{\psi}
\nc{\om}{\omega}
\nc{\Ga}{\Gamma}
\nc{\De}{\Delta}
\nc{\La}{\Lambda}
\nc{\Si}{\Sigma}
\nc{\Up}{\Upsilon}
\nc{\Ph}{\Phi}
\nc{\Ps}{\Psi}
\nc{\Om}{\Omega}
\nc{\ptl}{\partial}
\nc{\del}{\nabla}
\nc{\ov}{\overline}
\nc{\newcaption}[1]{\centerline{\parbox{15cm}{\caption{#1}}}}
\nc{\us}{U(1)$_S$}
\nc{\Rg}{$R_{\gamma\gamma}$}

\def\beq{\begin{equation}}
\def\eeq{\end{equation}}
\def\bmat{\begin{displaymath}}
\def\emat{\end{displaymath}}
\def\bear{\begin{eqnarray}}
\def\eear{\end{eqnarray}}
\def\ba{\begin{eqnarray}}
\def\ea{\end{eqnarray}}
\def\bery{\begin{array}}
\def\ery{\end{array}}
\def\bit{\begin{itemize}}
\def\eit{\end{itemize}}
\def\ben{\begin{enumerate}}
\def\een{\end{enumerate}}
\def\btab{\begin{tabular}}
\def\etab{\end{tabular}}
\def\btbl{\begin{table}}
\def\etbl{\end{table}}
\def\bfig{\begin{figure}[tb]}
\def\efig{\end{figure}}
\def\bpic{\begin{picture}}
\def\epic{\end{picture}}

\def\ga{\mathrel{\raise.3ex\hbox{$>$\kern-.75em\lower1ex\hbox{$\sim$}}}}
\def\la{\mathrel{\raise.3ex\hbox{$<$\kern-.75em\lower1ex\hbox{$\sim$}}}}
\def\gappeq{\mathrel{\rlap {\raise.5ex\hbox{$>$}}
{\lower.5ex\hbox{$\sim$}}}}
\def\lappeq{\mathrel{\rlap{\raise.5ex\hbox{$<$}}
{\lower.5ex\hbox{$\sim$}}}}

\def\gyr{{\rm \, G\kern-0.125em yr}}
\def\mev{{\rm \, Me\kern-0.125em V}}
\def\gev{{\rm \, Ge\kern-0.125em V}}
\def\tev{{\rm \, Te\kern-0.125em V}}

\def\lsim{\mathrel{\rlap{\lower4pt\hbox{\hskip1pt$\sim$}}
    \raise1pt\hbox{$<$}}}                
\def\gsim{\mathrel{\rlap{\lower4pt\hbox{\hskip1pt$\sim$}}
    \raise1pt\hbox{$>$}}}                

\newcommand{\hef}{\ensuremath{{}^4\mathrm{He}}}
\newcommand{\het}{\ensuremath{{}^3\mathrm{He}}}
\newcommand{\lisx}{\ensuremath{{}^6\mathrm{Li}}}
\newcommand{\lisv}{\ensuremath{{}^7\mathrm{Li}}}
\newcommand{\bes}{\ensuremath{{}^7\mathrm{Be}}}
\newcommand{\deut}{\ensuremath{\mathrm{D}}}
\newcommand{\trit}{\ensuremath{\mathrm{T}}}

\newcommand{\hyd}{\ensuremath{\mathrm{H}}}

\newcommand{\keV}{\ensuremath{\mathrm{keV}}}
\newcommand{\MeV}{\ensuremath{\mathrm{MeV}}}
\newcommand{\GeV}{\ensuremath{\mathrm{GeV}}}

\renewcommand{\sec}{\ensuremath{\mathrm{s}}}

\woctitle{Dark Matter, Hadron Physics and Fusion Physics}
\begin{document}
\title{Nucleosynthesis constraints on the faint vector portal}
%
%

\author{Anthony Fradette\inst{1} \and
        Maxim Pospelov\inst{1,2} \and
        Josef Pradler \inst{3}\fnsep\thanks{\email{josef.pradler@oeaw.ac.at}} \and
        Adam Ritz\inst{1}
}

\institute{%
Department of Physics and Astronomy, University of Victoria, Victoria, BC V8P 5C2, Canada
\and
Perimeter Institute for Theoretical Physics, Waterloo, ON N2J 2W9, Canada
\and
Institute of High Energy Physics, Austrian Academy of Sciences, A-1050 Vienna, Austria
          }

\abstract{
  New Abelian $U(1)'$ gauge bosons $V_{\mu}$ can couple to the
  Standard Model through mixing of the associated field strength
  tensor $V_{\mu\nu}$ with the one from hypercharge,
  $F_{\mu\nu}^Y$. Here we consider early Universe sensitivity to this
  vector portal and show that the effective mixing parameter with the
  photon, $\kappa$, is being probed for vector masses in the GeV
  ballpark down to values $10^{-10}\lesssim \kappa \lesssim 10^{-14}$
  where no terrestrial probes exist.
  The ensuing constraints are based on a detailed calculation of the
  vector relic abundance and an in-depth analysis of relevant
  nucleosynthesis processes.
}
\maketitle
\section{Introduction}
\label{intro}

The origins of our Universe may well be rooted in inflation or
alternative cataclysmic scenarios that regard the very earliest
moments of existence.
However, despite the impressive success of observational cosmology
over the past decades, the earliest true \textit{direct} window into
the beginnings remain observations of light element abundances. They 
concern the epoch of primoridal nuclear transformations at cosmic
times $t\gtrsim 1\,\sec$.
The overall concordance of the Big Bang nucleosynthesis (BBN)
predictions with the observationally inferred primordial values is one
of the most impressive successes of modern day cosmology and particle
physics. Today, BBN is used as a toolbox to put models of new physics
to a stringent test~\cite{Pospelov:2010hj}, whenever they predict some
interference with the the standard processes in the observable sector
at $t\gtrsim 1\,\sec$.
 
Under the assumption of a canonical sequence of cosmological events,
the Universe emerged from inflation and baryogenesis much prior to
BBN. Such sequence then allows one to put stringent constraints on
very weakly interacting sectors of new physics beyond the Standard
Model (SM).
The kinetic mixing of a new $U(1)'$ vector $V_\mu$ with hypercharge
$F^Y_{\mu\nu} V^{\mu\nu}$ is of particular interest as the mixing with
the photon leads to numerous experimental consequences and much
attention was devoted to this vector portal in recent
years~\cite{Essig:2013lka}. Below the electroweak scale, the coupling
of $V$ to the SM is essentially given by its mixing with the
photon~\cite{Holdom:1985ag},
\begin{align}
  {\cal L}_{\rm V}
  = -\frac{\kappa}{2} F_{\mu\nu} V^{\mu\nu} = e \kappa V_\mu J^\mu_{\rm
  em}.
\end{align}
With $\kappa$ and $m_V$ being the only free parameters, the model
provides a simple, and technically natural prototype scenario for a
light, weakly interacting new particle sector. In the following we
will concentrate on a St\"uckelberg origin of $m_V$ that allows to
maintain gauge invariance in $U(1)'$ without complicating the
phenomenology by hidden Higgs particles $h'$; see
\textit{e.g.}~\cite{Batell:2009yf} for the phenomenology and
\cite{Pospelov:2010cw} for cosmological constraints on the latter
scenario.

The SM decay modes of $V$ are well known. When hadronic decays are
kinematically accessible, one can use experimental data on the
$R$-ratio to infer couplings to photons in the time-like direction,
and hence to determine the decay rate $\Gamma_V$ and all branching
ratios. Below the di-muon threshold and for $m_V>1\,\MeV $ the vector
$V$ decays to electron-positron pairs only, thereby setting its
principal lifetime,
\begin{align}
\label{eq:tauV}
  \tau_V \simeq  \frac{3 }{\kappa^2\alpha m_V }   = 
  270\,\sec \times  \frac{1\,{\rm GeV} }{ m_V} \left( \frac{10^{-12}}{\kappa} \right)^2,
\end{align}
where $\alpha$ is electromagnetic fine structure constant.
In the following the cosmological consequences of $U(1)'$ vectors with
masses in the MeV-GeV range, and lifetimes long enough for the decay
products to directly influence primordial nucleosynthesis are
explored. These vectors have a parametrically small coupling to the
electromagnetic current, and thus an extremely small production cross
section for $e^+e^-\to V\gamma$,
$ \sigma_{\rm prod} \sim \kappa^2 \pi \alpha^2 / s \sim 10^{-54}\,
{\rm cm}^2 $
where we took $\sqrt{s}= 200$~MeV and $\kappa = 10^{-12}$ from above.
Such small couplings render these vector states completely
undetectable in terrestrial particle physics experiments.

Despite the tiny production cross section, any charged SM state that
is populated in appreciable number in the early Universe at
temperature $T\sim m_V$ may yet emit $V$. With the above ballpark
numbers in~(\ref{eq:tauV}), parametric estimates suggest that an
amount of MeV/baryon may be stored in $V$-particles.  Followed by late
decays back to SM states, visible energy is therefore being injected
into the primordial plasma at levels that are probed by BBN. The early
Universe is therefore likely to be the only ``laboratory'' where such
\textit{very} dark photons with $\kappa \sim 10^{-12}$ and smaller are
being probed. 
Here we report on the detailed analysis performed in~\cite{Fradette:2014sza}
where also CMB limits on even later decays were considered.  Previous
partial discussions of cosmological signatures of decaying dark
photons may also be found in \cite{Redondo:2008ec,Pospelov:2010cw}.

In the following we assume no other light states $\chi$ that are
charged under $U(1)'$. Therefore, there are no decays
$V\to \chi \bar\chi$ that potentially drain visible modes and thereby
ameliorate the derived limits from BBN.
Using some recent insight about the in-medium production of dark
vectors \cite{An:2013yfc,An:2013yua} (see also~\cite{Redondo:2013lna})
we first discuss the production of dark vectors in the next section
and explore constraints on $V$-decays into SM in
Sec.~\ref{sec:impact-bbn} before concluding with
Sec.~\ref{sec:conclusions}.

\section{Abundance prior to decay}
\label{sec-1}

\begin{figure}[tb]
\begin{center}
\includegraphics[width=.35\textwidth]{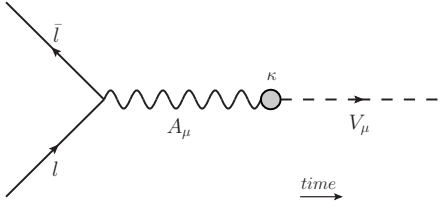}
\caption{Dark photon inverse decays are the leading contribution to
  sub-Hubble production rates in the calculation of the $V$ relic
  density.}
\label{fig:coall}
\end{center}
\end{figure}

The relic abundance of weakly coupled dark photons prior to their
decay is obtained through a calculation of the leakage from the
observable sector to the hidden sector with sub-Hubble rates,
$\Gamma_{\rm prod}/H \ll 1$. This ``freeze-in'' process is dominated
by inverse decays of $V$, through the coalescence of $e^\pm$,
$\mu^\pm$\dots, and, similarly, through hadronic
contributions; see the illustration in Fig.~\ref{fig:coall}.

The Boltzmann equation for the total number density of $V$ takes the form
\begin{align}
\label{Beq}
\dot{n}_V +3Hn_V = C  . 
\end{align}
The right hand side is the collision integral and provided that $V$
never reaches thermal equilibrium, it gives the number of $V$ states
emitted per unit volume and unit time.
In the Maxwell-Boltzmann approximation for the SM distribution
functions, and in the limit that only electrons coalesce, the
integration can be done analytically~\cite{Pospelov:2010cw,Fradette:2014sza}
(see also \cite{Redondo:2008ec})
\begin{align}
  C \simeq \frac{3}{2\pi^2} \Gamma_{V\to e\bar{e}} m_V^2 T K_1(m_V/T).
\end{align}
Here, $\Gamma_{V\to e\bar{e}}= \kappa^2\alpha m_V/3$ is the decay
width of $V$ to electrons, up to corrections $(m_e/m_V)^2$ and $K_1$
is the modified Bessel function of the second kind.
In terms of the number of $V$ particles normalized to entropy density,
$Y_V = n_V/s$, the cosmic time integral in (\ref{Beq}) for the final
``freeze-in'' abundance can be performed explicitly,
\begin{align}
\label{Ysimple}
Y_{V}^{(e)} = \frac{9}{4\pi} \frac{m_V^3\Gamma_{V\to e\bar{e}} }{(Hs)_{T=m_V}} .
\end{align}

While further leptonic production channels are easy to be included
into~(\ref{Ysimple}), hadronic production channels require assumptions
about the primordial hadron gas and the strength of interaction with
photons, denoted by $\alpha_{\rm eff}$.
At temperatures above the QCD confinement scale $T_c\sim 200\,\MeV$
light quarks are deconfined and individual quark contributions can be
added to $Y_{V}$ in a straightforward manner.  Below $T_c$ one may use
a free gas of mesons as an approximation to the hadronic
(non-baryonic) particle content in the early Universe.  The production
via inverse charged pion and kaon decays
$\{\pi^+\pi^- ,K^+K^-\} \to V$ can then be included using a scalar QED
model with effective coupling strengths like
$\alpha_{\rm eff}^{\pi\pi}(m_V) = \kappa^2 \alpha^{\pi\pi} (\sqrt{s} =
m_V)$
where $\alpha^{\pi\pi} $ is extracted from BaBar cross section
measurements of
$e^+e^- \to \gamma^* \to \pi^+\pi^-(\gamma)$~\cite{Lees:2012cj}, and
similarly for charged kaons~\cite{Lees:2013gzt}.

Finally, there is a possibility of resonant production of $V$ by
virtue of the thermal bath. Such in-medium effects may be cast into an
effective mixing angle,
\begin{align}
 \label{eq:keff}
  \kappa_{T,L} =  \frac{ \kappa }{|1 - \Pi_{T,L}/m_V^2|},
\end{align}
with $\Pi_{T,L}$ being the transverse ($T$) and longitudinal ($L$)
photon polarization functions in the primordial, isotropic plasma.
The expressions for $\Pi_{T,L}$ can \textit{e.g.}~be found
in~\cite{Braaten:1993jw}; the longitudinal polarization
function~\cite{An:2013yfc} used here is,
$\Pi_L^{\rm here} = m_V^2/(\omega^2-m_V^2)
\Pi_L^\text{Ref.\,\cite{Braaten:1993jw}}$
and $\omega $ is the (dark) photon energy. Equation~(\ref{eq:keff})
informs us about the condition of resonant dark photon production,
\begin{align}
\label{eq:resonance}
 \real \Pi_{T,L}(\omega, T_{r,T,L}) = m_V^2.
\end{align} 
The condition depends on temperature $T$ as $\Pi_{T,L}$ are
proportional to the plasma frequency, $\omega_P(T)$. Most importantly,
the resonance temperature $T_{r,T,L}(\omega)$ as a function of
frequency $\omega$ is parametrically larger than $m_V$ with a minimum
frequency at which the resonance can happen, 
\begin{align}
\label{Tmin}
T_{\rm r, min}  = m_V \left[ \frac{3}{2\pi\alpha}\right]^{1/2} \simeq 8 \,m_V.
\end{align}
Thus resonances occur at parametrically larger temperatures (by
$\alpha^{-1/2}$) than $m_V$, for which $H(T)$ is significantly larger
than at $T\simeq m_V$ at which the $V$ freeze-in production has its
biggest contribution.  Therefore, resonant contributions to $Y_V$ do
not alter the picture drastically  though numerically they may
constitute as much as 30\%.

After production, the momentum of $V$ redhifts quickly so that at the
time of decay the energy of $V$ is to good approximation given by the
rest mass, $E_V = m_V$.  The decay deposits this energy into leptons,
hadrons, and hadronic resonances.  The energy prior to decay that is
stored per baryon is therefore given by
\begin{align}
\label{Eyield} 
E_{\rm p.b.} = m_V Y_V \frac{s_0}{n_{b,0}},
\end{align}
where $n_{b,0}/s_0 = 0.9\times 10^{-10}$ is the baryon-to-entropy
ratio today.  Equipped with $E_{\rm p.b.}$ as a function of $m_V$ and
$\kappa$ following the detailed calculations of the $V$ ``freeze in''
abundance in \cite{Fradette:2014sza}, we may now explore its consequences for
BBN.

\section{\boldmath$V$-decays during BBN}
\label{sec:impact-bbn}

Primordial nucleosynthesis predictions are affected for dark photon
decays with cosmic lifetime $t\gtrsim 1\,\sec$ or larger.  Ensuing
constraints are then governed by a combination of lifetime and
abundance, both being complementary with respect to the vector mass:
$\tau_V$ ($Y_V$) decreases (increases) with larger $m_V$. From this
one expects constraints as localized islands in those parameters where
the epoch of primordial nucleosynthesis exhibits its greatest
sensitivity.

\subsection{Major effects and treatment}
\label{sec:major-effects}

The effects on BBN are understood by considering 
electromagnetic and hadronic energy injection separately.  Prior to
decay, the $V$ abundance relative to baryons is substantial,
$n_V/n_b \lesssim 10^{8}$ for $\tau_V < 1\,\sec$, and the decays of $V$
inject electrons, muons, and mesons in numbers larger than baryons.

Dark photon decays with $m_V \leq 2 m_{\pi^{\pm}} = 279\,\MeV$ result
exclusively in injection of electromagnetic energy, because
$V\to e^+ e^-, \mu^+\mu^-$ are the only kinematically accessible
modes.  Muons typically decay before interacting, and
electron-positron pairs are quickly thermalized by interactions with
background photons.  The resulting electromagnetic cascade with
spectrum $f_{\gamma}(E_{\gamma})$ entails a large number of
non-thermal photons that may then spall light elements.

Importantly, the spectrum has a relatively sharp cut-off for energies
above the $e^{\pm}$ pair-creation threshold,
$E_{\mathrm{pair}} \simeq m_e^2/(22T)$.  Photons with
$E_{\gamma} > E_{\rm pair}$ are being dissipated before they interact
with nuclei, and to good approximation $f_{\gamma}(E_{\gamma}) = 0$
for $E_{\gamma}>E_{\mathrm{pair}}$. Photons with
$E_{\gamma} < E_{\rm pair}$, however, undergo slower degradation
processes and may interact with the light elements before being
thermalized.  Equating $E_{\mathrm{pair}}$ against the
photo-destruction thresholds (in brackets below) yields the
temperature and thereby the cosmic time $t_{\rm ph}$ of biggest impact
for a spallation channel:
\begin{equation*}
  t_{\rm ph} \simeq 
\left\{ \begin{array}{ll@{}r}
2\times 10^4\sec ,&  \ \bes+\gamma\to\het+\hef& (1.59\,\MeV),\\
5\times 10^4\sec  ,&  \ \deut+\gamma\to n+ p & (2.22\,\MeV),\\
4\times 10^6\sec ,& \ \hef+\gamma\to \het/\trit+n/p& (20\,\MeV),
\end{array}\right .
\end{equation*}

The spallation rate of species $N$ with number density $n_N$ is given
by
\begin{align}
\label{eq:boltz}
  \Gamma_{\mathrm{ph}} (T) = 2 n_N \int_{E_{\mathrm{thr}}}^{E_{\mathrm{max}}}
  dE_{\gamma} \,  f_{\gamma}(E_\gamma) \sigma_{\gamma + N
    \to X}(E_\gamma),
\end{align}
where $\sigma_{\gamma + N \to X}(E_\gamma)$ is the photo-dissociation
cross section for $\gamma + N \to X$ with threshold
$E_{\mathrm{thr}}$. The factor of two accounts for the two independent
cascades that form in a back-to-back decay of $V$ at rest, each with a
maximum energy of
$E_{\mathrm{max}} = \max \left\{ E_{\mathrm{pair}}, E_{\mathrm{inj}}/2
\right\}$.
All spallation reactions listed in~\cite{Cyburt:2002uv} are taken into
account in the numerical analysis.  We note in passing that neutrino
injection from muon decay constitute only minor corrections to the
photon-induced processes listed above~\cite{Pospelov:2010cw}.

For vector masses above the di-pion threshold,
$m_V > 2 m_{\pi^{\pm}}$, hadronic modes are accessible in
the decay of $V$ and the effects on BBN are more intricate. In the
hadronic decay of $V$ only $\pi^{\pm}$, $K^{\pm}$, and $K_L$, with
lifetimes $\tau \sim 10^{-8}\,\sec$, and (anti-)nucleons have a chance
to undergo a strong interaction reaction before decaying by
themselves.

Before deuterium formation at $T \simeq 100\,\keV$, only charge
exchange reactions on nucleons, such as $\pi^- + p \to \pi^0 + n $,
are possible. They change the $n/p$ ratio and thereby most
prominently the primordial $\hef$ value. 
After the deuterium bottleneck---once light elements have
formed---charge exchange creates ``extra neutrons'' on top of the
residual and declining neutron abundance. In addition, absorption with
subsequent destruction of light elements such as
$\pi^- + \hef \to T + n$ is now operative.
Spallation of \hef\ may also have a secondary consequence: the
production of mass-3 elements with non-thermal kinetic energy may
induce reactions of the sort $\trit +\hef_{\rm bg}\to \lisx + n$.  In
the numerical analysis, these processes as well as secondary
populations of $\pi^{\pm}$ from kaon decays, and hyperon producing
channels from reactions of kaons on nucleons and nuclei are being
accounted for.
Furthermore, in our analysis, we restrict ourselves to reactions at
threshold, with charged pions and kaons being thermalized before
reacting on light elements; such approximation generally results in
more conservative constraints. A detailed quantitative discussion of
incomplete stopping can be found in~\cite{Pospelov:2010cw}.

\begin{figure}[tb]
\begin{center}
\includegraphics[width=0.95\columnwidth]{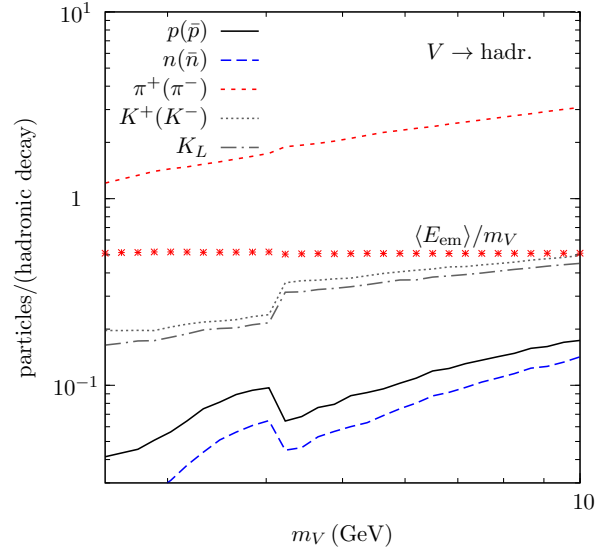}
\caption{Pythia simulation on the average number of particles per $V$
  decay and on the injected electromagnetic energy after all particles
  have decayed or annihilated to electrons and photons.  About one
  third of the energy is carried away by neutrinos. Narrow hadronic
  resonances are neglected.  }
\label{fig:pythia}
\end{center}
\end{figure}

Finally, baryon pairs are produced in the $V$-decay for
$m_V\gtrsim 2\,\GeV$.  Final state nucleons $\bar n$ and $\bar p$ will
preferentially annihilate on protons with an annihilation cross
section $\langle \sigma_{\rm ann}v\rangle \sim m_{\pi^{\pm}}^{-2}$.
The injection of $n\bar n $ then results in one net $p\to n$
conversion with associated energy injection of $m_p+m_n$.
Annihilation on neutrons with similar cross section is also possible
and $p\bar p $ injection results in one net $n\to p$
conversion. Assuming equal cross sections, the relative efficiency is
$p/(n+p) $ and $n/(n+p) $, respectively.

\begin{figure}[tb]
  \includegraphics[width=0.95\columnwidth]{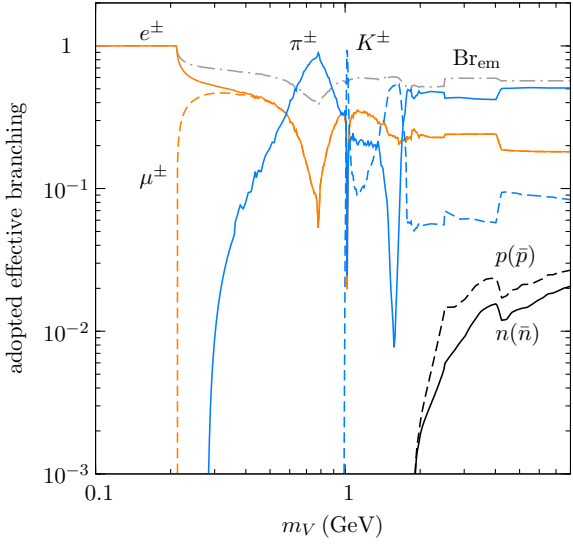}
  \caption{Final state branching ratios of long-lived mesons and other
    relevant decay products.  BaBar measurements of the
    $e^{\pm }\to\pi^{\pm}$ and $e^{\pm }\to K^{\pm}$ cross sections up
    to $m_V = 1.8\,\GeV$ are stiched together with a Pythia simulation
    starting at $m_V \geq 2.5\,\GeV$; a branching to $K_L$ was
    neglected and the fraction of $m_V$ that is ultimately being
    converted to electromagnetic energy is labeled~Br$_{\rm em}$. }
\label{fig:Vdec}
\end{figure}

At threshold, the rate for neutron injection can be inferred from a
measurement of the $e^+ e^- \to n \bar n $ cross section,
$\sigma_{e^+ e^- \to n \bar n }\sim 1\,\rm{nb}
$~\cite{Antonelli:1998fv}.
Given a total hadronic cross section
$ \sigma_{e^+ e^- \to \rm had } \sim 50\,\rm{nb} $ at this energy, the
branching fraction to a neutron-antineutron pair is $\sim 2\%$.
Away from the di-nucleon threshold, with multi-pion(kaon) production
and decays to hyperons and baryonic resonances being prevalent,
$V$-decays may be simulated using Pythia.
The ultimate yield of $\pi^{\pm}, K^{\pm}$, $K_L$, and nucleons prior
to their decay is shown in Fig.~\ref{fig:pythia}; dots depict the
average electromagnetic energy that is injected after all particles
have decayed to electrons and photons; $e^+$ have been annihilated on
$e^-$. The rest of the decay-energy is carried away by neutrinos.
At lower energies, decay events are eventually dominated by two body
decays. Above the di-pion (di-kaon) threshold, we use BaBar
measurements of the $e^{\pm }\to\pi^{\pm}$ and $e^{\pm }\to K^{\pm}$
cross section until an reported energy of
$\sqrt{s} = m_V = 1.8\,\GeV$. Relevant ultimate branching ratios are
shown in Fig.~\ref{fig:Vdec}; the effects of $K_L$ are, for
simplicity, neglected in our BBN anaslysis.

A more detailed discussion along with a list of all included reactions
can be found in the original paper~\cite{Fradette:2014sza} as well as in the
preceding work~\cite{Pospelov:2010cw}. Numerical results were obtained
by usage of a Boltzmann code that is based on
Ref.~\cite{Kawano:1992ua}, with significant improvements and updates
as detailed in~\cite{Pospelov:2010cw}. Standard BBN yields are found
to be in agreement with~\cite{Cyburt:2008kw} when using a baryon
asymmetry of $\eta_{b}= 6.2\times 10^{-10}$ and a neutron lifetime of
$\tau_n=885.7\,$s.

\subsection{Light element observations}
\label{sec:light-elem-obs}

BBN sensitivity is attained by the observational inferrence of light
element abundances and their estimated error bar. Here we briefly
discuss those observations that form the basis of our obtained regions
of interest.

The most abundant element after hydrogen is helium. Its mass fraction
$Y_p$ is inferred from extragalactic HII regions, and values in the
range
\begin{align}
\label{eq:he4obs}
  0.24\leq Y_p \leq 0.26
\end{align}
have been reported over the years. Owing to potential systematic
uncertainties~\cite{Izotov:2010ca,Aver:2013wba} we adopt
(\ref{eq:he4obs}) as the cosmologically viable range.

Among recent developments, the precision determination of D/H from
high redshift quasar absorption systems stands
out~\cite{Pettini:2012ph,Cooke:2013cba}.  Error bars have reduced by a
factor of five in comparison to previously available
determinations. The weighted mean now reads~\cite{Cooke:2013cba},
\begin{align}
\label{eq:D}
  {\rm D/H} = (2.53 \pm 0.04) \times 10^{-5} . 
\end{align}
D astration on dust grains is, however, a potential source of
systematic uncertainty, and values as high as $4\times 10^{-5}$ have
also been reported~\cite{Burles:1997fa,Kirkman:2003uv}. In light of
this, we adopt an upper limit of,
\begin{align}\label{eq:Dhigh} {\rm D/H} < 3\times 10^{-5} .
\end{align}
as well. Finally, producing too little D/H yields a robust limit
because no known astrophysical sources of this fragile light element
exist. We therefore either use the nominal lower $2\sigma$-limit from
(\ref{eq:D}) or require (robustly),
\begin{align}
\label{eq:hetconstr}
  \het/\deut < 1 
\end{align}
instead. The latter value is derived form solar system
observations~\cite{1993oeep.book.....P}.

Finally, and with much smaller abundance, the primordial value of
$\lisv/\hyd$~\cite{Spite:1982dd}, is lower than the lithium yield from
standard BBN by a factor of 3-5,
$\lisv/\hyd=(5.24^{+0.71}_{-0.67})\times
10^{-10}$~\cite{Cyburt:2008kw}.
We consider lithium being in concordance with observations if BBN
predictions yield
\begin{align}
   10^{-10}<\lisv/\hyd<2.5\times 10^{-10} . 
\end{align}
While new physics may be at the heart of the lithium problem, we
caution that astrophysical depletion mechanisms may also play their
part in solution to this long-standing puzzle;
see~\cite{Fields:2011zzb} for a recent review.

\begin{figure}[tb]
\begin{center}
\includegraphics[width=1.03\columnwidth]{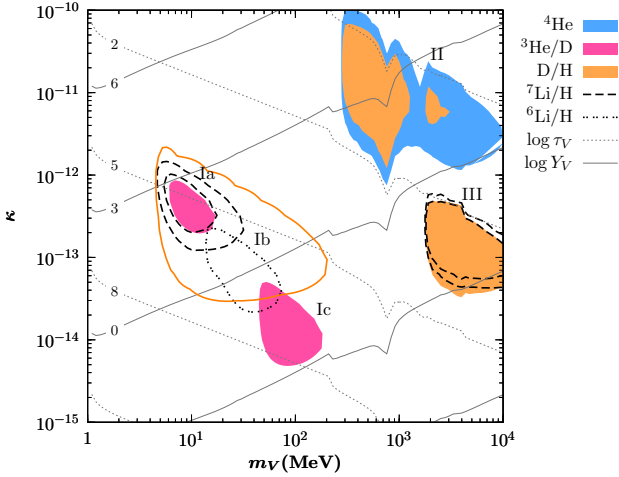}
\caption{Vector mass $m_V $ and kinetic mixing parameter $\kappa$
  parameter space with BBN sensitivity.  Diagonal gray contours depict
  $\tau_V$ (solid) or $n_V/n_b$ prior to decay (dotted). Shaded
  regions are excluded by observations as labeled.  The solid
  (orange) closed line is a $2\sigma$ constraint from underproduction
  of D/H derived from (\ref{eq:D}). Dashed black lines show
  decreasing levels of $\lisv/\hyd$, $4\times 10^{-10}$ and
  $3\times 10^{-10}$, from outer cirlces to the inner ones,
  respectively. Along the dotted line $\lisx/\hyd = 10^{-12}$
  signifying an two orders magnitude enhanced $\lisx $ yield, that is,
  however, not yet constrained by observations.}
\label{fig:bbn}
\end{center} 
\end{figure}

\subsection{Results}
\label{sec:results}

Our results from $V$-decays and their effect on BBN are presented in
the $m_V,\kappa$ parameter space in Fig.~\ref{fig:bbn}. Contours of
constant lifetime, $\tau_V$ and freeze-in abundance $n_V/n_b$ are
shown by the diagonal solid and dotted lines, respectively.  The
regions labeled I-III are in conflict with observations as detailed in
the previous section.

In regions I, $V$ decays to $e^+ e^-$  result in electromagnetic
energy injection.
Region Ia ($\tau_V\sim 10^5\,\sec$) is marked by a destruction of
\bes\ and D. The $\lisv/\hyd$ abundance is reduced to
$4\times 10^{-10}$ and $3\times 10^{-10}$ from the outside to the
inside, respectively. However, cosmologically favored smaller
$\lisv/\hyd$ abundances are challenged by$\het/\deut<1$ (pink shaded
region). Using~(\ref{eq:D}), lower $\lisv/\hyd$ values are
excluded by the nominal $2\sigma$ lower limit on $\deut/\hyd$ as
depicted by the solid closed line.
Region Ib is additionally marked by spallation of $\lisv$ and $\bes$
from non-thermal photons. This results in direct production of
$\lisx/\hyd> 10^{-12}$---values yet too low for being observationally
constrained at the moment.
Finally, Region Ic with $\tau_V \sim 10^7\,\sec$ is marked by $\hef$
dissociation and net creation of $\het/\deut$ ruling out this
parameter region. Secondary production of $\lisx$ is not efficient
enough to yield an additional limit.

In region~II, $\tau_V < 100\, \sec$ and $V$ decays before the end of
the D-bottleneck ($T\sim 100\,\keV$).  Injection of pions, kaons, and
nucleons, results in anomalous $n\leftrightarrow p$ inter-conversion.
The consequence is an elevated $n/p$-ratio and therefore enhanced D
and $\hef$ yields. The low-lifetime/high-abundance region~II is
correspondingly disfavored by $Y_p\leq 0.26$ and
$\deut/\hyd \leq 3\times 10^{-5}$.

Finally, region III is marked by the production of ``extra neutrons''
at $t\sim 10^3\,\sec$ from $V\to n \bar n$ and from charge exchange of
$\pi^-$ on protons, $\pi^- p\to n\pi^0 $ or $\pi^- p\to n \gamma$. In
addition, hyperon production by ``s-quark'' exchange of $K^-$ on
protons may also result in extra neutrons. With it comes a path that
may deplete lithium, $\bes + n \to \lisv + p ,$ followed by
$\lisv + p \to \hef + \hef . $
With a reduced Coulomb barrier, $\lisv$ is more susceptible to proton
burning in the second step and the declining $\lisv$ trend is depicted
by the dashed lines in Fig.~\ref{fig:bbn}. Most of the extra neutrons,
however, end up being captured by protons and the associated D/H
constraint~(\ref{eq:Dhigh}) is given by the orange region.

\section{Conclusions}
\label{sec:conclusions}

The kinetic mixing of a new $U(1)'$ gauge group with the Standard
Model $U(1)$ factors of hypercharge and, below the electroweak scale,
of electromagnetism is one of the few portals to the hidden sector
with renormalizable couplings. The associated gauge boson $V$ is often
called a ``dark photon'' and in this manuscript we have reported the
cosmological limits from BBN as they have been derived
in~\cite{Fradette:2014sza}. BBN sensitivity reaches photon kinetic
mixing parameters of $\kappa \sim 10^{-14}$ for
$1\,\MeV \leq m_V \lesssim 10\,\GeV$, unchallenged from terrestrial
dark photon searches, see, \textit{e.g.}~the works and
presentations~\cite{Beacham:2014vna,Celentano:2014wya,Battaglieri:2014qoa,Curciarello:2014nxa,Merkel:2014yya,Raggi:2014zpa,Essig:2010xa}

The presented limits are based on a thermal abundance of $V$ and a
standard cosmological history of the
Universe---\textit{i.e.}~uneventful until $V$-decay---with reheat
temperatures in excess of $m_V$. Additional contributions to the
$V$-abundance such as from an initial $V$-condensate after inflation
may only strengthen the derived bounds.
The latter source of primordial $V$-particles is particularly
interesting in the context of smaller $V$-masses. Below the
di-electron threshold (not considered in this work), $V$ has a
naturally long lifetime with $V\to 3\gamma$ being the only decay mode.
Therefore $V$ can even be a dark matter
candidate~\cite{Pospelov:2008jk,Redondo:2008ec} and ensuing
constraints on the photoelectric absorption of $V$ on atoms in dark
matter detectors have started to receive attention only very
recently~\cite{Pospelov:2008jk, Arisaka:2012pb, Abe:2014zcd,dddpdm}. 

\subsection*{Acknowledgements}
The speaker thanks S.~Eidelman for the invitation to the workshop.
JP is supported by the New Frontiers program by the Austrian Academy
of Sciences.

\end{document}